\documentstyle[epsfig,preprint,aps]{revtex}           
\begin{document}       
\tighten

\def\bea{\begin{eqnarray}}          
\def\eea{\end{eqnarray}}          
\def\beas{\begin{eqnarray*}}          
\def\eeas{\end{eqnarray*}}          
\def\nn{\nonumber}          
\def\ni{\noindent}          
\def\G{\Gamma}          
\def\d{\delta}          
\def\l{\lambda} 
\def\L{\Lambda}         
\def\g{\gamma}          
\def\m{\mu}          
\def\n{\nu}          
\def\s{\sigma}          
\def\tt{\theta}          
\def\b{\beta}          
\def\a{\alpha}          
\def\f{\phi}          
\def\fh{\phi}          
\def\y{\psi}          
\def\z{\zeta}          
\def\p{\pi}          
\def\e{\epsilon}          
\def\ve{\varepsilon}          
\def\cl{{\cal L}}          
\def\cv{{\cal V}}          
\def\cz{{\cal Z}} 
\def\co{{\cal O}}          
\def\pl{\partial}          
\def\ov{\over}          
\def\~{\tilde}          
\def\rar{\rightarrow}          
\def\lar{\leftarrow}          
\def\lrar{\leftrightarrow}          
\def\rra{\longrightarrow}          
\def\lla{\longleftarrow}          
\def\8{\infty} 
\def\h{\hbar} 
\def\lg{\langle}
\def\rg{\rangle}

\def\ag{\biggl({m^2_\f\ov 4\p\m^2}\biggr)} 
\def\agg{\biggl({m^2_\f\ov \m^2}\biggr)}       
\newcommand{\fr}{\frac}


\title{Renormalization group improvement of the 
effective potential in massive $\phi^4$ theory}
          
\author{J.~-M. Chung\footnote                  
  {Electronic address: chung@ctpa03.mit.edu}}
\address{Center for Theoretical Physics,
Massachusetts Institute of Technology,\\
Cambridge, Massachusetts 02139}

\author{B.~K. Chung\footnote                  
  {Electronic address: bkchung@nms.kyunghee.ac.kr}}               
\address{Research Institute for Basic Sciences                   
and Department of Physics\\ Kyung Hee University, Seoul  130-701, Korea\\
{~}} 
       
\date{MIT-CTP-2862,~~~~ May 1999}                  
\maketitle              
\draft              
\begin{abstract}           
\indent           
Using the method of renormalization group, 
we improve the two-loop effective potential of the massive $\phi^4$ theory
to obtain the next-next-to-leading logarithm correction  
in the $\overline{\rm MS}$ scheme. Our result well reproduces the 
next-next-to-leading logarithm parts of the ordinary loop expansion 
result known up to the four-loop order. 
\end{abstract}              
                   
\section{Introduction}             
The groundbreaking paper \cite{gl} of Gell-Mann and Low was in a large part 
directed to the problem of improving perturbation theory, i.e., to the problem
of using the ideas of the renormalization group and the results of perturbation
theory to a given order to say something about the next order of perturbation
theory. The method of the renormalization group improvement of perturbation
theory can be applied to the computation of the Green's functions and other 
predictions of Feynman diagram perturbation theory including the effective 
potential.

The effective potential in quantum field theory is a convenient tool
in probing the vacuum structure of the theory. The usual way of computing
the effective potential is a loop expansion \cite{cw}, for which an elegant 
method called the field shift method was developed by Jackiw \cite{jk}. It 
has been recognized for a long time that ordinary loop-wise perturbation 
expansions of important physical quantities are not only restricted to small 
values of the couplings but are often rendered useless by the occurrence of the 
large logarithms. Renormalization group resummation of these logarithms 
is then crucial to establish a region of validity for the perturbative 
results.

Renormalization-group-improved effective potentials were first considered in 
the context of massless models by Coleman and Weinberg \cite{cw}. 
In the massive case it has been demonstrated that this treatment also works 
provided one takes into account the running of the vacuum energy (or 
cosmological constant) \cite{br,bkmn,fjse,bkmn2}. In the papers of 
Ref.~\cite{bkmn2}, the multiscale problems were studied. 

The purpose of this paper is to improve the effective potential of the 
massive $\phi^4$ theory through the second-to-leading logarithm, i.e.,
next-next-to-leading logarithm, order
in the modified minimal subtraction ($\overline{\rm MS}$) scheme.
We compare the structures of the loop expansion and leading-logarithm series 
expansion of the effective potential in Sec.~II. 
In Sec.~III, the running parameters are determined through the three-loop 
order from their evolution equations with three-loop renormalization group
functions and the result of the next-next-to-leading logarithm
improvement is given. The final section is devoted to the concluding remarks.

\section{Loop Expansions vs. Leading-Logarithm Series Expansion}
The Lagrangian of the massive $\f^4$ theory is given as
\bea
\cl&=&{1\ov 2}(\pl \f)^2-{m^2\ov 2} \f^2-{\l\ov 4!}\f^4-\L\nn\\
&+&{\d Z\ov 2}(\pl \f)^2-{\d m^2\ov 2} \f^2-{\d\l\ov 4!}\f^4-\d\L\;.
\label{lg}
\eea
Here $\d Z$, $\d m^2$, and $\d \l$ are the so-called counterterms of the 
wave function, mass, and coupling constant, respectively. Their values 
are known up to the five-loop order for the massive $O(N)$ $\f^4$ theory 
in the $\overline{\rm MS}$ scheme from the renormalization of the two- and
four-point (one-particle-irreducible) Green's functions \cite{sf,sft}.
The constant $\L$, usually called the cosmological constant term,
is included in the above Lagrangian, Eq.~(\ref{lg}).
The corresponding counterterm is calculated up to 
the five-loop order for the massive $O(N)$ $\f^4$ theory in the 
$\overline{\rm MS}$  scheme \cite{ks}.
Such a term becomes relevant to the $\f$-dependent terms of the effective 
potential \cite{bkmn}. We note that these renormalization counterterms
can be determined either in the renormalization of the 1PI 
Green's functions $\G^{(0)}$, $\G^{(2)}$, and $\G^{(4)}$, or in the
renormalization of the effective potential $V$. In particular,  
the vacuum energy renormalization constant up to three-loop order is given
as 
\beas
\d \L={m^4\ov 2(4\p)^2\e}+{\l m^4\ov 2(4\p)^4\e^2}+{\l^2 m^4\ov (4\p)^6}
\biggl({5\ov 6\e^3}-{5\ov 18\e^2}+{1\ov 48\e}\biggr)\;.
\eeas

The effective potential of the theory defined by Eq.~(\ref{lg}) up to 
two-loop order in the $\overline{\rm MS}$ scheme \cite{fj,jm} is given 
as\footnote{The two-loop effective potential 
for the massive $O(N)$ $\f^4$ theory in the $\overline{\rm MS}$ scheme
and the three-loop effective potential for the single-component massive 
$\f^4$ theory were calculated in Ref.~\cite{fj} and Ref.~\cite{jm}
respectively. In Ref.~\cite{jm}, there is no difficulty
in obtaining the $\overline{\rm MS}$ result because at the 
intermediate stage of calculation, finite constants of the 
renormalization counterterms in dimensional regularization scheme
are kept as symbols, not as numerical values.} 
\bea
V&=&V^{(0)}+V^{(1)}+V^{(2)}\;,\nn\\
V^{(0)}&=&{m^2 \f^2\ov 2}+{\l \f^4\ov 4!}+\L\;,\nn\\
V^{(1)}&=&{\l\ov (4\p)^2}\biggl[-{3m^4\ov 8\l}-{3 m^2 \f^2\ov 8}
-{3\l\f^4\ov 32}
+\biggl\{{m^4\ov 4\l}+{m^2 \f^2\ov 4}
+{\l\f^4\ov 16}\biggr\}\ln\agg\biggr]\;,\nn\\
V^{(2)}&=&{\l^2\ov (4\p)^4}\biggl[{m^4\ov 8\l}
+m^2\f^2\biggl( {3\ov 4}+{A\ov 4}\biggr)
+\l\f^4\biggl({11\ov 32}+{A\ov 8}\biggr)\nn\\
&-&\biggl\{{m^4\ov 4\l}+{3m^2\f^2\ov 4}
+{5\l\f^4\ov 16}\biggr\}\ln\agg
+\biggl\{{m^4\ov 8\l}+
{m^2 \f^2\ov 4}+{3\l\f^4\ov 32}\biggr\}\ln^2\agg\biggr]\;,
\label{v012}
\eea
where $m_\f^2\equiv m^2+\l\f^2/2$ and  $A$ is a constant,
$A=-1.171953619\cdots$.

Now let us compare the structures of the loop expansion and the 
leading-logarithm series expansion of the effective potential. 
In the usual loop expansion, the $l$-loop quantum correction
to the effective potential has the following general structure 
\cite{br}: 
\bea
V^{(l)}(\f,\l,x,y)&=&
\l^{l+1}\f^4\sum_{m=0}^{l-1}\sum_{n=0}^la_{lmn}x^{m-2}y^n\;,
\label{le}
\eea
where
\bea
x\equiv{1\ov 1+2m^2/(\l\f^2)}\;,~~~~~~~~y\equiv \ln{m_\f^2\ov \m^2}\;.
\eea
Introducing a function $G_n^{(l)}$ as follows:
\beas
G_n^{(l)}(\f,\l,x)\equiv 
(4\p)^{2l}\l\f^4\sum_{m=0}^{l-1}a_{lm(l-n)}x^{m-2}\;,
\eeas
Eq.~(\ref{le}) can be rewritten as
\bea
V^{(l)}(\f,\l,x,y)={\l^l\ov(4\p)^{2l}}\sum_{n=0}^lG_n^{(l)}(\f,\l,x)y^{l-n}\;.
\label{le1}
\eea
In terms of the original parameters, this equation can be expressed as
\bea
V^{(l)}(\f,\l,m^2,\m^2)={\l^l\ov(4\p)^{2l}}\sum_{n=0}^l G_n^{(l)}(\f,\l,m^2)
\ln^{l-n}\biggl({m_\f^2\ov \m^2}\biggr)\;. \label{leo}
\eea 
For example, the four-loop contribution, $V^{(4)}$, takes the following form
\bea
V^{(4)}&=&{\l^4\ov (4\p)^8} \biggl[A_1{m^4\ov \l}
+A_2m^2\f^2+A_3\l\f^4+A_4{\l\f^4\ov 1+2m^2/(\l\f^2)}\nn\\ 
&+&\biggl\{B_1{m^4\ov \l}
+B_2m^2\f^2+B_3\l\f^4+B_4{\l\f^4\ov 1+2m^2/(\l\f^2)}\biggr\}\ln\agg\nn\\ 
&+&\biggl\{C_1{m^4\ov \l}
+C_2m^2\f^2+C_3\l\f^4+C_4{\l\f^4\ov 1+2m^2/(\l\f^2)}\biggr\}\ln^2\agg\nn\\
&+&\biggl\{D_1{m^4\ov \l}
+D_2m^2\f^2+D_3\l\f^4+D_4{\l\f^4\ov 1+2m^2/(\l\f^2)}\biggr\}\ln^3\agg\nn\\
&+&\biggl\{E_1{m^4\ov \l}
+E_2m^2\f^2+E_3\l\f^4+E_4{\l\f^4\ov 1+2m^2/(\l\f^2)}\biggr\}\ln^4\agg\biggr]\;.
\label{v4}
\eea
In addition to the $l$-loop correction, Eq.~(\ref{le1}) or Eq.~(\ref{leo}),
there is a tree-level potential, called classical potential, which can be 
expressed as
\bea
V^{(0)}=\f^4\biggl(a\l+{b\l\ov x}\biggr)+\L\equiv G_0^0(\f,\l,x,\L)
= G_0^0(\f,\l,m^2,\L)\;.\label{v0}
\eea
The (complete) effective potential is given as the sum of all loop 
corrections and the above tree-level potential $V^{(0)}$:
\bea
V=\sum_{l=0}^\8{\l^l\ov (4\p)^{2l}}\sum_{n=0}^l G_n^{(l)}\ln^{l-n}(m_\f^2/\m^2)\;.
\label{veff-loop}
\eea
Notice that the zero-point energy level is set to be  
\bea
V(\f=0)=\L+{\l^l\ov (4\p)^{2l}}\sum_{l=1}^\8 G_l^{(l)}(0,\l,m^2)=-\G^{(0)}\;.
\eea
This choice was taken correctly by the authors of Ref.~\cite{bkmn}.

Rearranging the summation order in Eq.~(\ref{veff-loop}), 
we can reexpress the effective potential as follows:
\bea
V&=&\sum_{l=0}^\8\sum_{n=l}^\8 G_l^{(n)}[\l^n/(4p)^{2n}]
\ln^{n-l}(m_\f^2/\m^2)\nn\\
&=&\sum_{l=0}^\8[\l^l/(4p)^{2l}]\sum_{n=l}^\8 G_l^{(n)}\~{y}^{n-l}\nn\\
&\equiv&\sum_{l=0}^\8 [\l^l/(4\p)^{2l}] F_l(\f,\l,x,\~{y})\nn\\
&\equiv&\sum_{l=0}^\8 V^{(\!(l)\!)}\;,\label{veff-log}
\eea
where $\~{y}\equiv\l y/(4\p)^2=(\l/(4\p)^2)\ln (m_\f^2/\m^2)$.
This form of expansion, Eq.~(\ref{veff-log}), in powers of $\l$,
was first derived by Kastening \cite{br}. The term proportional to
$\l^l$ in $V$ is referred to as an $l$th-to-leading logarithm 
term \cite{bkmn}. The functions $F_0$, $F_1$, ... correspond to the leading, 
next-to-leading, ... logarithm terms, respectively.\footnote{
The function 
$F_l$ defined in Eq.~(\ref{veff-log}) is related to the Kastening's function 
$f_l$ as follows:
\beas
F_l=(4\p)^{2l}(\l\f^4)f_{l+1}\;.
\eeas
}
This concept of leading-logarithm
series expansion is shown schematically in Fig.~1.

\section{Running Parameters and the Next-Next-To-Leading
Logarithm Improvement}               
In his remarkable paper \cite{br}, Kastening has introduced two methods ---
the power series method and the differential equations method --- 
to improve the effective potential in massive $\f^4$ theory.
Though we readily get a closed-form expression for $F_0$ from the obtained
recursion relations for $a_{lmn}$, the power series method is too much 
complicated to extend it beyond leading-logarithm correction.   
The differential equations method, which is a smarter one,
does not rely on being able to sum up a
power series involving coefficients gotten through complicated recursion
relations. Instead, we obtain the recursion relations for the Kastening's
function $f_l$ themselves from the renormalization group equation.

We use here the method of characteristics to improve the effective potential
of the massive $\f^4$ theory up to the next-next-to-leading 
logarithm correction for the first time: 
though there is an elegant method of Ref.~\cite{bkmn},
in which we use only $G_l^{(l)}$ as boundary functions, 
we follow closely the method used in Sec.~II of the second paper in 
Ref.~\cite{bkmn2}.  

Since we assume the effective potential $V(=\sum_{l=0}^\8 V^{(l)}=
\sum_{l=0}^\8V^{(\!(l)\!)}$) is independent of the 
renormalization scale $\m$ for the fixed values of the bare parameters, 
the effective potential $V(\m,\l,m^2,\f,\L)$ obeys the renormalization group 
equation
\bea
\biggl[\m{\pl\ov \pl\m}+\b_{\l}{\pl\ov \pl\l}+\g_m m^2{\pl\ov \pl m^2}
-\g_\f\f{\pl\ov \pl\f}+\b_\L{\pl\ov\pl\L}\biggr]V(\m,\l,m^2,\f,\L)=0\;.
\label{rge}
\eea
The various $\b$ and $\g$ functions ($\b_{\l}$, $\g_m$, $\g_\f$, and $\b_\L$) 
introduced in the above equation are 
known up to the five-loop order \cite{sf,sft,ks}. Here we quote these
values up to the three-loop order: 
\bea
\b_\l&=&{3\l^2\h\ov (4\p)^2}-{17\l^3\h^2\ov 3(4\p)^4}
+{\l^4\h^3\ov (4\p)^6}\biggl({145\ov 8}+12\z(3)\biggr)\nn\\
&\equiv&\b_1\l^2\h+\b_2\l^3\h^2+\b_3\l^4\h^3\;,\nn\\
\g_m&=&{\l\h\ov (4\p)^2}-{5\l^2\h^2\ov 6(4\p)^4}+{7\l^3\h^3\ov 2(4\p)^6}
\nn\\
&\equiv&\g_{m1}\l\h+\g_{m2}\l^2\h^2+\g_{m3}\l^3\h^3\;,\nn\\
\g_\f&=&{\l\h\ov (4\p)^2}\times 0+{\l^2\h^2\ov 12(4\p)^4}
-{\l^3\h^3\ov 16(4\p)^6}\nn\\
&\equiv&\g_1\l\h+\g_2\l^2\h^2+\g_3\l^3\h^3\;,\nn\\
\b_\L&=&{m^4\h\ov 2(4\p)^2}+{m^4\l\h^2\ov (4\p)^4}\times 0
+{m^4\l^2\h^3\ov 16(4\p)^6}\nn\\
&\equiv&m^4(\b_{\L1}\h+\b_{\L2}\l\h^2+\b_{\L3}\l^2\h^3)\;.
\label{rgfunc}
\eea
Note that in the above equation, $\h$ factors are inserted;
we will use the $\h$ factor as a counting parameter in the leading-logarithm 
series expansion.\footnote{In Eq.~(\ref{v012}), $\h$ factors are inserted, too,  
as follows:
\beas
V&=&V^{(0)}+\h V^{(1)}+\h^2 V^{(2)}+O(\h^3)\;.
\eeas 
} 
Applying the method of characteristics to Eq.~(\ref{rge}), we can write
the solution of Eq.~(\ref{rge}), $V(\m,\l,m^2,\f,\L)$ as follows:
\bea
V(\m,\l,m^2,\f,\L)=V(\bar{\mu},\bar{\l},\bar{m}^2,\bar{\f},\bar{\L})\;,
\label{char}
\eea
where the running parameters satisfy
\bea
&&\h{d\bar{\m}\ov dt}=\bar{\m}\;,~~~~
\h{d\bar{\l}\ov dt}=\b_\l(\bar{\l})\;,~~~~
\h{d\bar{m}^2\ov dt}=\g_m(\bar{\l})\bar{m}^2\;,\nn\\
&&\h{d\bar{\f}\ov dt}=-\g_\f(\bar{\l})\bar{\f}\;,~~~~
\h{d\bar{\L}\ov dt}=\b_\L(\bar{\l},\bar{m}^2)\;,\label{bar}
\eea
and at the boundary point, $t=0$, their values are given as $\bar{\m}(t=0)=\m$,
$\bar{m}^2(t=0)=m^2$, $\bar{\f}(t=0)=\f$, and $\bar{\L}(t=0)=\L$. 

The solution to $\bar{\m}$ differential equation is given as
\beas
\bar{\m}^2(t)=\m^2 \exp(2t/\h)\;.
\eeas
In order to solve $\bar{\l}$ differential 
equation, we try a perturbative solution by writing
\beas
{\bar \l}={\bar \l}^{\lg 0\rg}+\h{\bar \l}^{\lg 1\rg}
+\h^2{\bar \l}^{\lg 2\rg}+O(\h^3)\;,
\eeas
with the boundary conditions
$\bar{\l}^{\lg 0\rg}(0)=\l$, ${\bar \l}^{\lg 1\rg}(0)={\bar \l}^{\lg 2\rg}(0)=0$.
Then, with $\b_\l$ in Eq.~(\ref{rgfunc}), the equation we want to solve splits
into three equations within the desired order:
\beas
{d\bar{\l}^{\lg 0\rg}\ov dt}&=&\b_1\bar{\l}^{\lg 0\rg 2}\;,\nn\\
{d\bar{\l}^{\lg 1\rg}\ov dt}&=&2\b_1\bar{\l}^{\lg 0\rg}\bar{\l}^{\lg 1\rg}
+\b_2\bar{\l}^{\lg 0\rg 3}\;,\nn\\
{d\bar{\l}^{\lg 2\rg}\ov dt}&=&2\b_1\bar{\l}^{\lg 0\rg}\bar{\l}^{\lg 2\rg}
+\b_1\bar{\l}^{\lg 1\rg 2}
+3\b_2\bar{\l}^{\lg 0\rg 2}\bar{\l}^{\lg 1\rg}+
\b_3\bar{\l}^{\lg 0\rg 4}\;.
\eeas
Solutions to these equations are obtained as 
\bea
\bar{\l}^{\lg 0\rg}&=&{\l\ov T}\;,\nn\\
\bar{\l}^{\lg 1\rg}&=&-{\b_2\l^2\ov \b_1 T^2}\ln T\;,\nn\\
\bar{\l}^{\lg 2\rg}&=&{\l^3\ov T^2}\biggl[\biggl(-{\b_2^2\ov \b_1^2}
+{\b_3\ov \b_1}\biggr)[T^{-1}-1]
-{\b_2^2\ov \b_1^2}{\ln T\ov T}+{\b_2^2\ov \b_1^2}{\ln^2 T\ov T}\biggr]\;,
\label{bls}
\eea
where $T\equiv 1-\b_1\l t$.
Similarly, we write $\bar{m}^2$ as 
\beas
{\bar m}^2={\bar m}^{2\lg 0\rg}+\h{\bar m}^{2\lg 1\rg}
+\h^2{\bar m}^{2\lg 2\rg}+O(\h^3)\;,
\eeas
and with $\g_m$ in Eq.~(\ref{rgfunc}),
obtain splitted equations:          
\beas
{d\bar{m}^{2\lg 0\rg}\ov dt}&=&\g_{m1}\bar{\l}^{\lg 0\rg}\bar{m}^{2\lg 0\rg}
\;,\nn\\
{d\bar{m}^{2\lg 1\rg}\ov dt}&=&\g_{m1}\bar{\l}^{\lg 0\rg}\bar{m}^{2\lg 1\rg}
+\g_{m1}\bar{\l}^{\lg 1\rg}\bar{m}^{2\lg 0\rg}
+\g_{m2}\bar{\l}^{\lg 0\rg 2}\bar{m}^{2\lg 0\rg}\;,\nn\\
{d\bar{m}^{2\lg 2\rg}\ov dt}&=&\g_{m1}\bar{\l}^{\lg 0\rg}\bar{m}^{2\lg 2\rg}
+\g_{m1}\bar{\l}^{\lg 2\rg}\bar{m}^{2\lg 0\rg}
+\g_{m1}\bar{\l}^{\lg 1\rg}\bar{m}^{2\lg 1\rg}
+2\g_{m2}\bar{\l}^{\lg 0\rg}\bar{\l}^{\lg 1\rg}\bar{m}^{2\lg 0\rg}\nn\\
&+&\g_{m2}\bar{\l}^{\lg 0\rg 2}\bar{m}^{2\lg 1\rg}
+\g_{m3}\bar{\l}^{\lg 0\rg 3}\bar{m}^{2\lg 0\rg}\;.
\eeas
With the $\bar{\l}$ solutions, Eq.~(\ref{bls}), together with the boundary
conditions $\bar{m}^{2\lg 0\rg}(0)=m^2$, 
${\bar m}^{2\lg 1\rg}(0)={\bar m}^{2\lg 2\rg}(0)=0$,
we obtain the following $\bar{m}^2$ solutions: 
\bea
\bar{m}^{2\lg 0\rg}&=&{m^2\ov T^{\g_{m1}/\b_1}}\;,\nn\\
\bar{m}^{2\lg 1\rg}&=&{\l m^2\ov T^{\g_{m1}/\b_1}}\biggl[\biggl(
-{\b_2\g_{m1}\ov  \b_1^2}+{\g_{m2}\ov \b_1}\biggr)[T^{-1}-1]
-{\b_2\g_{m1}\ov \b_1^2}{\ln T\ov T}\biggr]\;,\nn\\
\bar{m}^{2\lg 2\rg}&=&{\l^2 m^2\ov T^{\g_{m1}/\b_1}}\biggl[\biggl( 
-{\b_2^2\g_{m1}\ov 2\b_1^3}+{\b_3\g_{m1}\ov 2\b_1^2}
+{\b_2^2\g_{m1}^2\ov 2\b_1^4}-{\b_2\g_{m2}\ov 2\b_1^2}\nn\\
&-&{\b_2\g_{m1}\g_{m2}\ov \b_1^3}+{\g_{m2}^2\ov 2\b_1^2}
+{\g_{m3}\ov 2\b_1}\biggr)[T^{-2}-1]\nn\\
&+&\biggl({\b_2^2\g_{m1}\ov \b_1^3}
-{\b_3\g_{m1}\ov \b_1^2}-{\b_2^2\g_{m1}^2\ov \b_1^4}
+{2\b_2\g_{m1}\g_{m2}\ov \b_1^3}-{\g_{m2}^2\ov\b_1^2})[T^{-1}-1]\nn\\
&+&\biggl(-{\b_2^2\g_{m1}^2\ov \b_1^4}+{\b_2\g_{m1}\g_{m2}\ov \b_1^3}\biggr)
{\ln T\ov T}+\biggl({\b_2^2\g_{m1}^2\ov \b_1^4}
-{\b_2\g_{m2}\ov \b_1^2}-{\b_2\g_{m1}\g_{m2}\ov \b_1^3}\biggr)
{\ln T\ov T^2}\nn\\
&+&\biggl({\b_2^2\g_{m1}\ov 2\b_1^3}+{\b_2^2\g_{m1}^2\ov 2\b_1^4}
\biggr){\ln^2 T\ov T^2}\biggr]\;.\label{bms}
\eea
If we note that $\bar{\f}$ differential equation is of the same structure except
the minus sign on the right-hand side, one may readily read off
the perturbation solutions for 
${\bar \f}={\bar \f}^{\lg 0\rg}+\h{\bar \f}^{\lg 1\rg}
+\h^2{\bar \f}^{\lg 2\rg}+O(\h^3)$ as follows:
\bea
\bar{\f}^{\lg 0\rg}&=&{\f\ov T^{-\g_{1}/\b_1}}\;,\nn\\
\bar{\f}^{\lg 1\rg}&=&{\l \f\ov T^{-\g_{1}/\b_1}}\biggl[
\biggl({\b_2\g_{1}\ov  \b_1^2}-{\g_{2}\ov \b_1}\biggr)[T^{-1}-1]
+{\b_2\g_{1}\ov  \b_1^2}{\ln T\ov T}\biggr]\;,\nn\\
\bar{\f}^{\lg 2\rg}&=&{\l^2 \f\ov T^{-\g_{1}/\b_1}}\biggl[ 
\biggl({\b_2^2\g_{1}\ov 2\b_1^3}-{\b_3\g_{1}\ov 2\b_1^2}
+{\b_2^2\g_{1}^2\ov 2\b_1^4}+{\b_2\g_{2}\ov 2\b_1^2}
-{\b_2\g_{1}\g_{2}\ov \b_1^3}+{\g_{2}^2\ov 2\b_1^2}
-{\g_{3}\ov 2\b_1}\biggr)[T^{-2}-1]\nn\\
&+&\biggl(-{\b_2^2\g_{1}\ov \b_1^3}
+{\b_3\g_{1}\ov \b_1^2}-{\b_2^2\g_1^2\ov \b_1^4}
+{2\b_2\g_{1}\g_{2}\ov \b_1^3}-{\g_{2}^2\ov\b_1^2}\biggr)[T^{-1}-1]\nn\\
&+&\biggl(-{\b_2^2\g_{1}^2\ov \b_1^4}+{\b_2\g_{1}\g_{2}\ov \b_1^3}\biggr)
{\ln T\ov T}
+\biggl({\b_2^2\g_{1}^2\ov \b_1^4}
+{\b_2\g_{2}\ov \b_1^2}-{\b_2\g_{1}\g_{2}\ov \b_1^3}\biggr)
{\ln T\ov T^2}\nn\\
&+&\biggl(-{\b_2^2\g_{1}\ov 2\b_1^3}+{\b_2^2\g_1^2\ov 2\b_1^4}
\biggr){\ln^2 T\ov T^2}\biggr]\;.\label{bfs}
\eea
Finally, we try the solution to the $\bar{\L}$ differential equation as 
\beas
{\bar \L}={\bar \L}^{\lg 0\rg}+\h{\bar \L}^{\lg 1\rg}
+\h^2{\bar \L}^{\lg 2\rg}+O(\h^3)\;.
\eeas
Then from the splitted equations
\beas
{d\bar{\L}^{\lg 0\rg}\ov dt}&=&\b_{\L1}\bar{m}^{2\lg 1\rg 2}\;,\nn\\
{d\bar{\L}^{\lg 1\rg}\ov dt}&=&2\b_{\L1}\bar{m}^{2\lg 0\rg}\bar{m}^{2\lg 1\rg}
+\b_{\L2}\bar{\l}^{\lg 0\rg}\bar{m}^{2\lg 0\rg 2}\;,\nn\\
{d\bar{\L}^{\lg 2\rg}\ov dt}&=&\b_{\L1}\bar{m}^{2\lg 1\rg 2}
+2\b_{\L1}\bar{m}^{2\lg 0\rg}\bar{m}^{2\lg 2\rg}
+\b_{\L2}\bar{\l}^{\lg 1\rg}\bar{m}^{2\lg 0\rg 2}\nn\\
&+&2\b_{\L2}\bar{\l}^{\lg 0\rg}\bar{m}^{2\lg 0\rg}\bar{m}^{2\lg 1\rg}
+\b_{\L3}\bar{\l}^{\lg 0\rg 2}\bar{m}^{2\lg 0\rg 2}\;,
\eeas
and the boundary conditions ${\bar \L}^{\lg 0\rg}=\L$,
${\bar \L}^{\lg 1\rg}={\bar \L}^{\lg 2\rg}=0$, we obtain
\bea
\bar{\L}^{\lg 0\rg}&=&\L
-{m^4\b_{\L1}\ov \l(\b_1-2\g_{m1})}[T^{1-2\g_{m1}/\b_1}-1]\;,\nn\\
\bar{\L}^{\lg 1\rg}&=&m
^4\biggl[{2\b_{\L1}\ov \b_1(\b_1-2\g_{m1})}\biggl(
-{\b_2\g_{m1}\ov \b_1}+\g_{m2}\biggr)
[T^{1-2\g_{m1}/\b_1}-1]\nn\\
&-&\biggl({\b_2\b_{\L1}\ov \b_1^2}+{\b_2\b_{\L1}\ov 2\b_1\g_{m1}}
-{\b_{\L1}\g_{m2}\ov \b_1\g_{m1}}-{\b_{\L2}\ov 2\g_{m1}}\biggr)
[T^{-2\g_{m1}/\b_1}-1]\nn\\
&-&{\b_{\L1}\b_2\ov \b_1^2}{\ln T\ov T^{2\g_{m1}/\b_1}}\biggr]\;,\nn\\
\bar{\L}^{\lg 2\rg}&=&\l m^4 \biggl[\biggl(
{\b_2\b_{\L2}\ov \b_1^2}
-{\b_3\b_{\L1}\ov \b_1^2}-{2\b_2^2\b_{\L1}\g_{m1}\ov \b_1^4}
+{4\b_2\b_{\L1}\g_{m2}\ov \b_1^3}\nn\\
&-&{\b_{\L2}\g_{m2}\ov \b_1\g_{m1}}+{\b_2\b_{\L1}\g_{m2}\ov \b_1^2\g_{m1}}
-{2\b_{\L1}\g_{m2}^2\ov \b_1^2\g_{m1}}\biggr)
[T^{-2\g_{m1}/b_1}-1]\nn\\
&+&{\b_{\L1}\ov \b_1(\b_1-2\g_{m1})}\biggl({\b_2^2\g_{m1}\ov \b_1^2}
-{\b_3\g_{m1}\ov \b_1}-{2\b_2^2\g_{m1}^2\ov \b_1^3}\nn\\
&-&{\b_2\g_{m2}\ov \b_1}+{4\b_2\g_{m1}\g_{m2}\ov \b_1^2}-{2\g_{m2}^2\ov \b_1}
+\g_{m3}\biggr)[T^{1-2\g_{m1}/\b_1}-1]\nn\\
&+&{1\ov \b_1+2\g_{m1}}\biggl(
\b_{\L3}-{\b_2\b_{\L2}\ov b_1}-{2\b_2\b_{\L2}\g_{m1}\ov \b_1^2}\nn\\
&+&{\b_2^2\b_{\L1}\g_{m1}\ov \b_1^3}
+{\b_3\b_{\L1}\g_{m1}\ov \b_1^2}
+{2\b_2^2\b_{\L1}\g_{m1}^2\ov \b_1^4}
+{2\b_{\L2}\g_{m2}\ov \b_1}\nn\\
&-&{3\b_2\b_{\L1}\g_{m2}\ov \b_1^2}
-{4\b_2\b_{\L1}\g_{m1}\g_{m2}\ov \b_1^3}
+{2\b_{\L1}\g_{m2}^2\ov \b_1^2}
+{\b_{\L1}\g_{m3}\ov \b_1}\biggr)[T^{-1-2\g_{m1}/\b_1}-1]\nn\\
&+&\biggl({2\b_2\b_{\L1}\g_{m2}\ov \b_1^3}
-{2\b_2^2\b_{\L1}\g_{m1}\ov \b_1^4}\biggr)
{\ln T\ov T^{2\g_{m1}/\b_1}}\nn\\
&+&\biggl(-{\b_2\b_{\L2}\ov \b_1^2}
+{2\b_2^2\b_{\L1}\g_{m1}\ov \b_1^4}
-{2\b_2\b_{\L1}\g_{m2}\ov \b_1^3}\biggr)
{\ln T\ov T^{1+2\g_{m1}/\b_1}}\nn\\
&+&{\b_2^2\b_{\L1}\g_{m1}\ov \b_1^4}{\ln^2 T\ov T^{1+2\g_{m1}/\b_1}} \biggr]\;.
\label{bLs}
\eea
  
If we insert the numerical values in Eq.~(\ref{rgfunc}) for the symbols
$\b_i$,$\g_{mi}$, $\g_i$, and $\b_{\L i}$ into 
Eqs.~(\ref{bls}) --- (\ref{bLs}),
we obtain the running parameters up to the 
lowest three orders: 
\bea
\bar{\l}(t)&=&\l\biggl(1-{3\l t\ov (4\p)^2}\biggr)^{-1}+{17\h\l^2\ov 9(4\p)^2}
\biggl(1-{3\l t\ov (4\p)^2}\biggr)^{-2}\ln\biggl(1-{3\l t\ov (4\p)^2}\biggr)\nn\\
&+&{\h^2\l^3\ov (4\p)^4}\biggl\{
-\biggl(1-{3\l t\ov (4\p)^2}\biggr)^{-2}\biggl[{1603\ov 648}+4\z(3)\biggr]
+\biggl(1-{3\l t\ov (4\p)^2}\biggr)^{-3}
\biggl[{1603\ov 648}+4\z(3)\nn\\
&-&{289\ov 81}\ln\biggl(1-{3\l t\ov (4\p)^2}\biggr)
+{289\ov 81}\ln^2\biggl(1-{3\l t\ov (4\p)^2}\biggr)\biggr]
\biggr\}+O(\h^3)\;,\nn\\
\bar{m}^2(t)&=&m^2\biggl(1-{3\l t\ov (4\p)^2}\biggr)^{-1/3}\nn\\
&+&{\h\l m^2\ov (4\p)^2}\biggl\{
-{19\ov 54}\biggl(1-{3\l t\ov (4\p)^2}\biggr)^{-1/3}
+\biggl(1-{3\l t\ov (4\p)^2}\biggr)^{-4/3}\biggl[{19\ov 54}+{17\ov 27}\ln
\biggl(1-{3\l t\ov (4\p)^2}\biggr)\biggr]\biggr\}\nn\\
&+&{\h^2\l^2 m^2\ov (4\p)^4}\biggl\{
\biggl(1-{3\l t\ov (4\p)^2}\biggr)^{-1/3}\biggl[{1787\ov 11664}+{2\z(3)\ov 3}
\biggr]\nn\\
&-&\biggl(1-{3\l t\ov (4\p)^2}\biggr)^{-4/3}\biggl[{5531\ov 5832}+{4\z(3)\ov 3}
+{323\ov 1458}\ln\biggl(1-{3\l t\ov (4\p)^2}\biggr)\biggr]\nn\\
&+&\biggl(1-{3\l t\ov (4\p)^2}\biggr)^{-7/3}\biggl[{9275\ov 11664}+{2\z(3)\ov 3}
-{221\ov 729}\ln\biggl(1-{3\l t\ov (4\p)^2}\biggr)
+{578\ov 729}\ln^2\biggl(1-{3\l t\ov (4\p)^2}\biggr)\biggr]
\biggr\}\nn\\
&&+O(\h^3)\;,\nn\\
\bar{\f}(t)&=&\f+{\h\l\f\ov 36(4\p)^2}\biggl\{
1-\biggl(1-{3\l t\ov (4\p)^2}\biggr)^{-1}\biggr\}
+{\h^2\l^2\f\ov (4\p)^4}\biggl\{{7\ov 432}\nn\\
&-&{1\ov 1296}\biggl(1-{3\l t\ov (4\p)^2}\biggr)^{-1}
-\biggl(1-{3\l t\ov (4\p)^2}\biggr)^{-2}
\biggl[{5\ov 324}+{17\ov 324}\ln\biggl(1-{3\l t\ov (4\p)^2}\biggr)\biggr]
\biggr\}+O(\h^3)\;,\nn\\
\bar{\L}(t)&=&\L+{m^4\ov 2\l}\biggl\{1-
\biggl(1-{3\l t\ov (4\p)^2}\biggr)^{1/3}\biggr\}\nn\\
&+&{\h m^4\ov (4\p)^2}\biggl\{
-1+{19\ov 54}\biggl(1-{3\l t\ov (4\p)^2}\biggr)^{1/3}
+\biggl(1-{3\l t\ov (4\p)^2}\biggr)^{-2/3}
\biggl[{35\ov 54}+{17\ov 54}\ln\biggl(1-{3\l t\ov (4\p)^2}\biggr)\biggr]\nn\\
&+&{\h^2\l m^4\ov (4\p)^4}\biggl\{{23\ov 30}+{6\z(3)\ov 5}
-\biggl(1-{3\l t\ov (4\p)^2}\biggr)^{1/3}\biggl[{2509\ov 11664}
+{2\z(3)\ov 3}\biggr]\nn\\
&-&\biggl(1-{3\l t\ov (4\p)^2}\biggr)^{-2/3}\biggl[
{10129\ov 11664}+{2\z(3)\ov 3}
+{323\ov 1458}\ln\biggl(1-{3\l t\ov (4\p)^2}\biggr)\biggr]\nn\\
&+&\biggl(1-{3\l t\ov (4\p)^2}\biggr)^{-5/3}\biggl[
{9239\ov 29160}+{2\z(3)\ov 15}
+{323\ov 1458}\ln\biggl(1-{3\l t\ov (4\p)^2}\biggr)+
{289\ov 1458}\ln^2\biggl(1-{3\l t\ov (4\p)^2}\biggr)\biggr]\biggr\}\nn\\
&&+O(\h^3)\;.\label{rp}
\eea
If we choose $t$ as
\beas
t={\h\ov 2}\ln(m_\f^2/\m^2)\;,
\eeas
then $\bar{\m}^2(t)$ becomes
\bea
\bar{\m}^2(t)=m^2+{\l\f^2\ov 2}\;,\label{bm}
\eea
which is independent of $\m$. 
Substitute $\bar{\l}$, $\bar{m}^2$, $\bar{\f}$, and $\bar{\L}$ 
of Eq~(\ref{rp}), and $\bar{\m}^2$ of Eq.~(\ref{bm}) into the
right-hand side of Eq.~(\ref{char}) with $V$ as two-loop approximation 
in Eq.~(\ref{v012}). Then rearrange resulting terms in $\h$ order.
In this rearranging process, the running scale $t$ {\em should not} be
replaced with ${\h\ov 2}\ln(m_\f^2/\m^2)$. This is very important
for the correct collection of logarithms of various powers 
into a given leading-logarithm series order. 
By this rearrangement, we can write the effective potential $V$ as 
\beas
V=V^{(\!(0)\!)}(\f,\l,m^2,t,\L)+\h V^{(\!(1)\!)}(\f,\l,m^2,t)+
\h^2 V^{(\!(2)\!)}(\f,\l,m^2,t)+O(\h^3)\;,
\eeas
where
\bea
V^{(\!(0)\!)}&=&\L+{m^4\ov 2\l}(1-T^{1/3})
+{m^2\f^2\ov 2}T^{-1/3}+{\l\f^4\ov 24}T^{-1}\;,\nn\\
V^{(\!(1)\!)}&=&{\l\ov (4\p)^2}\biggl[{m^4\ov\l}\biggl\{-1+{19\ov 54}T^{1/3}
+T^{-2/3}\biggl({59\ov 216}+{17\ov 54}\ln T+{1\ov 4}
\ln S\biggr)\biggr\}\nn\\
&+&m^2\f^2\biggl\{-{4\ov 27}T^{-1/3}+T^{-4/3}\biggl(-{49\ov 216}
+{17\ov 54}\ln T +{1\ov 4}
\ln S\biggr)\biggr\}\nn\\
&+&\l\f^4\biggl\{{1\ov 216}T^{-1}+T^{-2}\biggl(
-{85\ov 864}+{17\ov 216}\ln T+{1\ov 16}
\ln S\biggr)\biggr\}\biggr]\;,\nn\\
V^{(\!(2)\!)}&=&{\l^2\ov (4\p)^4}\biggl[{m^4\ov\l}\biggl\{{23\ov 30}
+{6\z(3)\ov 5}-T^{1/3}\biggl({2509\ov 11664}+{2\z(3)\ov 3}\biggr)\nn\\
&-&T^{-2/3}\biggl({7051\ov 11664}+{2\z(3)\ov 3}+{323\ov 1458}\ln T
+{19\ov 108}\ln S\biggr)+T^{-5/3}\biggl({5189\ov 29160}
+{2\z(3)\ov 15}\nn\\
&-&{731\ov 2916}\ln T+{289\ov 1458}\ln^2 T-{2\ov 27}\ln S
+{17\ov 54}\ln T\ln S+{1\ov 8}\ln^2 S\biggr)\biggr\}\nn\\
&+&m^2\f^2\biggl\{T^{-1/3}\biggl({973\ov 11664}+{\z(3)\ov 3}\biggr)
-T^{-4/3}\biggl({4025\ov 11664}+{2\z(3)\ov 3}+{68\ov 729}\ln T
+{2\ov 27}\ln S\biggr)\nn\\
&+&T^{-7/3}\biggl({1475\ov 1458}+{A\ov 4}+{\z(3)\ov 3}
-{850\ov 729}\ln T+{289\ov 729}\ln^2 T-{73\ov 108}\ln S
+{17\ov 27}\ln T\ln S+{1\ov 4}\ln^2 S\biggr)\biggr\}\nn\\
&+&\l\f^4\biggl\{{5\ov 1728}T^{-1}+T^{-2}\biggl(
-{197\ov 1728}-{\z(3)\ov 6}+{17\ov 1944}\ln T
+{1\ov 144}\ln S\biggr)\nn\\
&+&T^{-3}\biggl({131\ov 288}+{A\ov 8}+{\z(3)\ov 6}
-{2023\ov 3888}\ln T+{289\ov 1944}\ln^2 T-{23\ov 72}\ln S
+{17\ov 72}\ln T\ln S+{3\ov 32}\ln^2 S\biggr)\biggr\}\nn\\
&+&{m^6\ov \l W}\biggl\{-{19\ov 216}T^{-1}+T^{-2}\biggl(
{19\ov 216}+{17\ov 108}\ln T\biggr)\biggr\}\nn\\
&+&{m^4\f^2\ov W}\biggl\{-{35\ov 432}T^{-5/3}
+T^{-8/3}\biggl({35\ov 432}+{85\ov 216}\ln T\biggr)\biggr\}\nn\\
&+&{\l m^2 \f^4\ov W}\biggl\{-{13\ov 864}T^{-7/3}
+T^{-10/3}\biggl({13\ov 864}+{119\ov 432}\ln T\biggr)\biggr\}\nn\\
&+&{\l^2 \f^6\ov W}\biggl\{{1\ov 576}T^{-3}
+T^{-4}\biggl(-{1\ov 576}+{17\ov 288}\ln T\biggr)\biggr\}\biggr]\;,
\label{result}
\eea
with 
\beas
T&\equiv&1-{3\l t\ov (4\p)^2}\;,\nn\\
W&\equiv&m^2T^{-1/3}+(\l\f^2/2)T^{-1}\;,\nn\\
S&\equiv&{W\ov m^2+\l\f^2/2}\;.
\eeas

\section{Concluding Remarks}               
In this paper, using the the method of renormalization group we have improved 
the two-loop effective potential for the (single-component) massive $\f^4$ theory
for the first time. In obtaining our result the various three-loop
renormalization group functions have been used. 

We first compare the existing result of lower-order calculations.
Our result $V^{(\!(0)\!)}$ and $V^{(\!(1)\!)}$ correspond to the Kastening's
functions $f_1$ and $f_2$, respectively. We compare them by subtracting one from 
the other:
\beas
&&V^{(\!(0)\!)}-\l\f^4 f_1=\L+{m^4\ov 2\l}\;,\nn\\
&&V^{(\!(1)\!)}-\l^2\f^4 f_2=-{m^4\ov (4\p)^2}\;.
\eeas
Only differences are $\f$-independent constant terms. 
These discrepancies between our result and 
Kastening's result are due to his not introducing a vacuum energy term in the
Lagrangian. He has made a peculiar {\it Ansatz}  
for it as a working means \cite{br}. 
Thus he used even the two-loop effective potential in obtaining
next-to-leading logarithm correction for fixing the coefficient ($b_2$) in
the {\it Ansatz}, contrary to the following general principle \cite{bkmn}: with
the $L$-loop effective potential and $(L+1)$-loop renormalization group
functions, we can obtain an renormalization-group-improved effective potential
which is exact up to the $L$th-to-leading logarithm order.  
It is remarkable that he has obtained  the correct $\f$-dependent terms from 
the ansatz. Other calculations of the leading-logarithm corrections,  
which exactly agree with our result $V^{(\!(0)\!)}$, can be found in 
Ref.~\cite{bkmn} and in the second paper of Ref.~\cite{bkmn2}.

In order to make the correctness check for $V^{(\!(2)\!)}$ 
richer, we add the three-loop
correction, $V^{(3)}$, which can be readily obtained from the Ref.\cite{jm},
to the two-loop effective potential of Eq.~(\ref{v012}):
\bea
V^{(3)}&=&{\l^3\ov (4\p)^6} \biggl[Q_1{m^4\ov \l}
+Q_2m^2\f^2+Q_3\l\f^4\nn\\ 
&+&\biggl\{{41m^4\ov 96\l}
+m^2\f^2\biggl( {371\ov 96}+{7A\ov 8}\biggr)
+\l\f^4\biggl({701\ov 384}+{9A\ov 16}
+{\z(3)\ov 4}\biggr)\biggr\}\ln\agg\nn\\ 
&-&\biggl\{{17m^4\ov 48\l}+{37m^2\f^2\ov 24}
+{143\l\f^4\ov 192}\biggr\}\ln^2\agg\nn\\ 
&+&\biggl\{{5m^4\ov 48\l}+{7m^2\f^2\ov 24}+
{9\l\f^4\ov 64}\biggr\}\ln^3\agg\biggr]\;,\label{v3}
\eea
where $Q_1$, $Q_2$, and $Q_3$ are constants and their
numerical values \cite{jmjmjm} are given as
\beas
Q_1=-0.5123\cdots\;, ~~~~~~~
Q_2=-1.8105\cdots\;, ~~~~~~~
Q_3=-0.9428\cdots\;.
\eeas
Further all the coefficients in Eq.~(\ref{v4}), except $A_1$,
$A_2$, $A_3$, and $A_4$ are determined \cite{jmjmjm}, 
from the fourth order part of Eq.~(\ref{rge}) in perturbation theory
\beas
\m{\pl V^{(4)}\ov \pl\m}&+&\l\biggl\{\b_1{\pl V^{(3)}\ov \pl\l}
+\g_{m1} m^2{\pl V^{(3)}\ov \pl m^2}
-\g_1 \f{\pl V^{(3)}\ov \pl\f}+\b_{\L1}{\pl V^{(3)}\ov\pl\L}\biggr\}\nn\\
&+&\l^2\biggl\{\b_2{\pl V^{(2)}\ov \pl\l}
+\g_{m2} m^2{\pl V^{(2)}\ov \pl m^2}
-\g_2 \f{\pl V^{(2)}\ov \pl\f}+\b_{\L2}{\pl V^{(2)}\ov\pl\L}\biggr\}\nn\\
&+&\l^3\biggl\{\b_3{\pl V^{(1)}\ov \pl\l}
+\g_{m3} m^2{\pl V^{(1)}\ov \pl m^2}
-\g_3 \f{\pl V^{(1)}\ov \pl\f}+\b_{\L3}{\pl V^{(1)}\ov\pl\L}\biggr\}\nn\\
&+&\l^4\biggl\{\b_4{\pl V^{(0)}\ov \pl\l}
+\g_{m4} m^2{\pl V^{(0)}\ov \pl m^2}
-\g_4 \f{\pl V^{(0)}\ov \pl\f}+\b_{\L4}{\pl V^{(0)}\ov\pl\L}\biggr\}=0\;, 
\eeas
as
\bea
B_1&=&-{295\ov 192}+4Q_1+{\z(3)\ov 4}\;,\nn\\
B_2&=&-{271\ov 32}-{53A\ov 48}+5Q_2-{15\z(3)\ov 8}
-{3\z(4)\ov 4}\;,\nn\\
B_3&=&-{1549\ov 768}-{35A\ov 96}+6Q_3-{9\z(3)\ov 4}
+{3\z(4)\ov 8}-{5\z(5)\ov 2}\;,\nn\\
B_4&=&{A\ov 8}+{\z(3)\ov 4}\;,\nn\\
C_1&=&{73\ov 48}\;,~~~~~~~
C_2={97\ov 8}+{35A\ov 16}+{3\z(3)\ov 4}\;,\nn\\
C_3&=&{583\ov 96}+{27A\ov 16}+{9\z(3)\ov 8}\;,~~~~~~~
C_4=-{1\ov 16}\;,\nn\\
D_1&=&-{55\ov 96}\;, ~~~~~~
D_2=-{277\ov 96}\;, ~~~~~~
D_3=-{201\ov 128}\;, ~~~~~~
D_4={1\ov 48}\;, \nn\\
E_1&=&{5\ov 48}\;,~~~~~~~
E_2={35\ov 96}\;, ~~~~~~~
E_3={27\ov 128}\;, ~~~~~~~
E_4=0\;.\label{bcde}
\eea
Our result of $V^{(\!(2)\!)}$ well reproduce
the next-next-to-leading logarithm parts of $V^{(2)}$ in Eq.~(\ref{v012}),
$V^{(3)}$ in Eq.~(\ref{v3}), and $V^{(4)}$ in Eq.~(\ref{v4}) and
Eq.~(\ref{bcde}) too, when it is expanded in power series of $t$,
as it should.

\acknowledgments               
This work was supported in part by Ministry of Education, Project number               
1998-015-D00073. J.~-M. C. was also supported in part by the Korea Science 
and Engineering Foundation, and in part by funds provided by the U.S.
Department of Energy (D.O.E.) under cooperative research agreement number
DF-FC02-94ER40818.


\begin{figure}           
  {\unitlength1cm           
   \epsfig{file=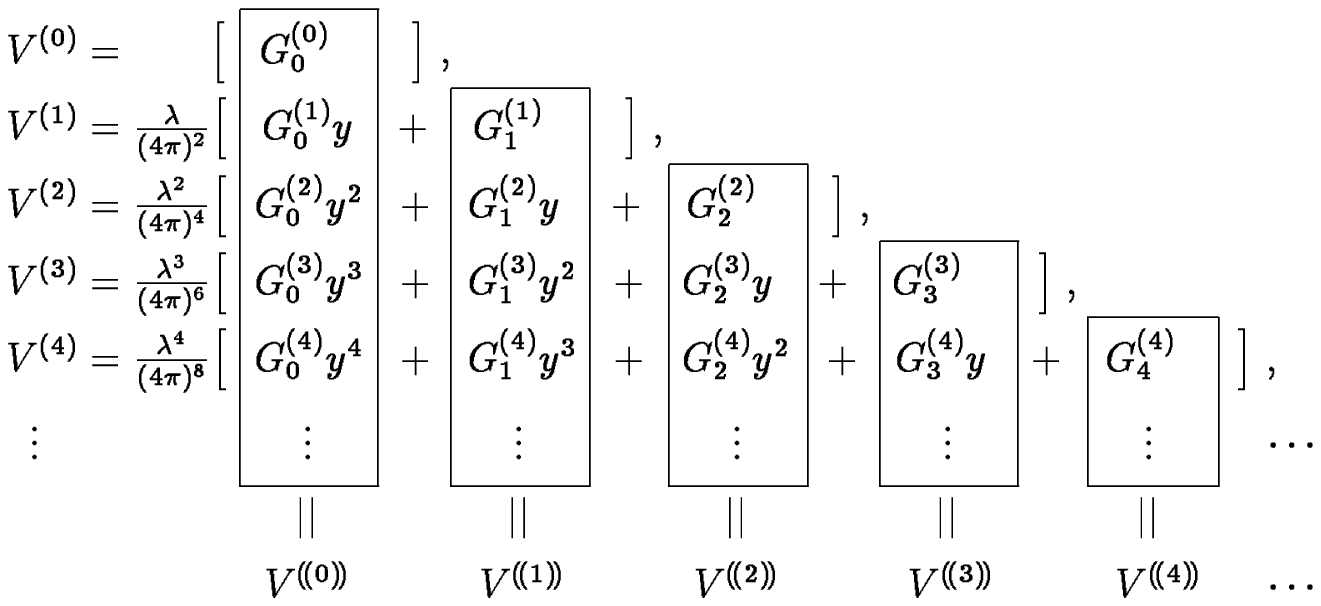,           
   bbllx=80pt,bblly=0pt,bburx=612pt,bbury=760pt,           
      rheight=8cm, rwidth=10cm,  clip=,angle=0} }             
\caption{The loop expansion and the leading-logarithm series expansion 
of the effective potential. It is understood that each vertical sum marked
by a box should be multiplied by the common factors [$\l^l/(4\p)^{2l}$]
in front of the horizontal sums to give the leading-logarithm series expansion; 
for example, $V^{(\!(0)\!)}=G_0^{(0)}+[\l/(4\p)^2]G_0^{(1)}y+
[\l^2/(4\p)^4]G_0^{(2)}y^2+...$, 
$V^{(\!(1)\!)}=[\l/(4\p)^2]G_1^{(1)}+
[\l^2/(4\p)^4]G_1^{(2)}y+[\l^3/(4\p)^6]G_1^{(3)}y^2+...$, etc.}
\end{figure}


\begin{thebibliography}{99}
\bibitem{gl}    
M. Gell-Mann and F. E. Low, Phys. Rev. {\bf 95}, 1300 (1954).                 

\bibitem{cw}       
S. Coleman and E. Weinberg, Phys. Rev. D {\bf 7}, 1888 (1973).

\bibitem{jk}                  
R. Jackiw, Phys. Rev. D {\bf 9}, 1686 (1974). 

\bibitem{br}                 
B. Kastening, Phys. Lett. {\bf B283}, 287 (1992). 

\bibitem{bkmn}                 
M. Bando, T. Kugo, N. Maekawa, and H. Nakano,
Phys. Lett. {\bf B301}, 83 (1993).

\bibitem{fjse}                 
C. Ford, D. R. T. Jones, P. W. Stephenson, and M. B. Einhorn,
Nucl. Phys. {\bf B395}, 17 (1993).

\bibitem{bkmn2}    
M. Bando, T. Kugo, N. Maekawa, and H. Nakano,
Prog. Theor. Phys. {\bf 90}, 405 (1993); 
C. Ford, Phys. Rev. D {\bf 50}, 7531 (1994); 
C. Ford and C. Wiesendanger,  
{\em ibid.}
{\bf 55}, 2202 (1997);
Phys. Lett. {\bf B398}, 342 (1997).

\bibitem{sf}     
H. Kleinert, J. Neu, V. Schulte-Frohlinde, K. G. Chetyrkin, and S. A.   
Larin, Phys. Lett. {\bf B272}, 39 (1991); {\bf B319}, 545(E) (1993). 
  
\bibitem{sft}
V. Schulte-Frohlinde, ``Renormalization of the $\Phi^4$-Theory'' (Thesis,
Freien Universit\"{a}t Berlin, 1996).

\bibitem{ks}                 
B. Kastening, Phys. Rev. D {\bf 54}, 3965 (1996); {\em ibid.} {\bf 57}, 
3567 (1998). 
     
\bibitem{fj}       
C. Ford and D. R. T. Jones, Phys. Lett. {\bf B274}, 409 (1992);        
{\bf B285}, 399(E) (1992). 

\bibitem{jm}               
J. -M. Chung and B. K. Chung, Phys. Rev. D {\bf 56}, 6508 (1997); 
{\em ibid.} {\bf 59} 109902 (E) (1999). 
   
\bibitem{jmjmjm}               
J. -M. Chung and B. K. Chung, in preparation.
\end{thebibliography}
\end{document}